\documentclass[aps,prl,twocolumn,superscriptaddress,showpacs]{revtex4-2}

\usepackage{caption}
\usepackage{subcaption}
\usepackage{graphicx}
\usepackage{amsmath}

\usepackage[colorlinks=true, citecolor=blue, linkcolor=blue, urlcolor=blue]{hyperref}

\usepackage{bm}
\usepackage{color}
\DeclareTextSymbol{\degre}{OT1}{23}
\usepackage{ulem}
\usepackage{siunitx}

\begin{document}

\title{Kondo scaling of $4f$-electron states and the Kondo singlet breakdown in heavy fermions}

\author{B. Tegomo Chiogo}
\email {bodry.tegomo\_chiogo@helmholtz-berlin.de}
\affiliation{Helmholtz-Zentrum Berlin für Materialien und Energie, Hahn-Meitner-Platz 1, Berlin D-14109, Germany}

\author{M. Tagliavini}
\affiliation{Heidelberg University, Institute for Theoretical Physics (ITP), Philosophenweg 19, 69120, Heidelberg, Germany}

\author{D. Wong}
\affiliation{Helmholtz-Zentrum Berlin für Materialien und Energie, Hahn-Meitner-Platz 1, Berlin D-14109, Germany}

\author{C. Schulz}
\affiliation{Helmholtz-Zentrum Berlin für Materialien und Energie, Hahn-Meitner-Platz 1, Berlin D-14109, Germany}

\author{V. Por\'ee}
\affiliation{Synchrotron SOLEIL, L'Orme des Merisiers, Saint-Aubin, Boîte Postale 48, F-91192 Gif-sur-Yvette, France}
\affiliation{Universit\'e de Rennes, CNRS, Institut des Sciences Chimiques de Rennes-UMR6226, 35042 Rennes, France}

\author{A. Nicolaou}
\affiliation{Synchrotron SOLEIL, L'Orme des Merisiers, Saint-Aubin, Boîte Postale 48, F-91192 Gif-sur-Yvette, France}

\author{R. Feyerherm}
\affiliation{Helmholtz-Zentrum Berlin für Materialien und Energie, Hahn-Meitner-Platz 1, Berlin D-14109, Germany}

\author{T. Schweitzer}
\affiliation{Universit\'e de Lorraine, CNRS, Institut Jean Lamour,
F-54000 Nancy,
France}

\author{T. Mazet}
\affiliation{Universit\'e de Lorraine, CNRS, Institut Jean Lamour,
F-54000 Nancy,
France}

\author{M. W. Haverkort}
\affiliation{Heidelberg University, Institute for Theoretical Physics (ITP), Philosophenweg 19, 69120, Heidelberg, Germany}

\author{A. Chainani}
\affiliation{National Synchrotron Radiation Research Center, Hsinchu Science Park, Hsinchu 30076, Taiwan}

\author{D. Malterre}
\affiliation{Universit\'e de Lorraine, CNRS, Institut Jean Lamour,
F-54000 Nancy,
France}

\author{K. Habicht}
\affiliation{Helmholtz-Zentrum Berlin für Materialien und Energie, Hahn-Meitner-Platz 1, Berlin D-14109, Germany}
\affiliation{Institut für Physik und Astronomie,
Universität Potsdam, D-14476 Potsdam, Germany}

\date{\today}

\begin{abstract}
The low-energy spin- and charge-sensitive thermodynamic properties of a broad range of strongly correlated 4f-electron systems follow Kondo scaling, with a characteristic Kondo temperature, $T_K$. While the theory is known for thermodynamic properties and high-energy spectroscopies of Kondo materials, the surface sensitivity of electron spectroscopy limits the extent to which Kondo scaling can be verified quantitatively. In this study, bulk-sensitive photon-in photon-out temperature-dependent 
resonant inelastic X-ray scattering (RIXS), in combination with single-impurity Anderson model (SIAM) calculations, is used to provide quantitative evidence of low- and high-energy Kondo scaling in CeSi$_2$. RIXS Ce M$_5$-edge spectra show a clear decrease in the  occupancy of the $f^0$ state as temperature increases accompanied by an increase of the spectral weight of the $f^1\underline L^1$ state, in good agreement with the SIAM calculations. The results demonstrate the breakdown of the Kondo singlet state, coupled with thermal occupation of the low-lying first-excited magnetic states. The RIXS data reveal a temperature evolution of the $f^n$ spectral weights, which is in stark contrast to that extracted from photoemission and inverse photoemission spectroscopies.
This study provides an accurate spectroscopic method to determine the Kondo energy $k_B$$T_K$ that is consistent with thermodynamic measurements, and highlights soft X-ray RIXS as a quantitative bulk probe of low- and high-energy-scale hybridization effects in strongly correlated materials.

\end{abstract}

\maketitle
%

Intermediate-valence compounds of Ce and Yb exhibit fascinating electronic and magnetic properties, such as heavy fermion formation, unconventional superconductivity, quantum phase transitions, magnetic field- and pressure-induced non-Fermi liquid behavior \cite{hewson1993, Coleman, Stewart,Sangphet2026}. The diversity of these phenomena results from a competition between the Ruderman-Kittel-Kasuya-Yosida (RKKY) magnetic exchange interaction and the many-body Kondo effect. This competition leads to different phases which are summarized in the well-known Doniach phase diagram \cite{DONIACH1977}. The Kondo interaction arises from the very large intra-atomic Coulomb interaction $U_{ff}$ in the $f$ shell, and the hybridization between localized $f$-electron states and delocalized conduction-band states, resulting in an antiferromagnetic exchange interaction between the local moments and the conduction electrons. At low hybridization, the RKKY interaction dominates, leading to a long-range magnetically ordered state, whereas at high hybridization, the ground state is a non-magnetic Kondo singlet.

 In the Kondo scenario relevant for a single-impurity (or an assembly of independent impurities), theoretical studies have clarified that physical quantities and especially the occupation number $n_f$ of the $f$ shell in these materials are expected to exhibit a universal scaling as a function of $T/T_{K} $\cite{Bickers1987}. Here, $n_f$ is obtained from the contributions of $f^{0}$, $f^{1}\underline L$, and $f^{2}\underline L^2$ basis states, which characterize the ground state and satisfactorily reproduce the experimental spectra with appropriate electronic parameters. The Kondo temperature $T_{K}$ is a material-dependent quantity, and it marks a crossover from a local moment regime at high temperature to a Fermi liquid regime at low temperature in which the local moments are screened by the conduction electrons. Thus, a precise knowledge of the microscopic electronic parameters that define the electronic structure, and especially $T_{K}$, are crucial to understanding the emergent exotic properties.

The Kondo energy scale $k_BT_K$, associated with the spin degree of freedom, can be determined from thermodynamic measurements such as magnetic susceptibility and specific heat. In contrast, the charge degrees of freedom of the $4f$ shell, characterized by the $4f$ level energy $\epsilon_f$ and the on-site Coulomb interaction $U_{ff}$, can be obtained from high-energy spectroscopies such as X-ray absorption spectroscopy (XAS), valence-band photoemission spectroscopy (PES), and inverse photoemission spectroscopy (IPES) \cite{Fuggle1983, Malterre1992, Malterre1992_2, Gunnarsson1983, Grioni1997, Dallera2004, sekiyama2000}. The energy resolution of these techniques is much larger than $k_B T_K$, and they can not directly resolve the Kondo energy scale. However, $T_K$ can be extracted from the scaling behavior of the spectral weight as a function of $T/T_K$. However, these spectroscopies are surface-sensitive, and surface effects can obscure the results leading to inconsistent electronic structure parameters compared those extracted from bulk thermodynamic measurements.
Temperature-dependent PES and IPES measurements have revealed characteristic changes in the spectral weight of the $4f$ configurations with increasing temperature.
More specifically, the $f^{0}$ structure is temperature independent, while the $f^{2}$ spectral weight increases, accompanied by a reduction of the Kondo resonance associated with the $f^{1}$ configuration.

Recent advances in resonant inelastic x-ray scattering (RIXS) instrumentation have turned this photon-in photon-out technique into a powerful probe of the bulk electronic structure of strongly correlated materials \cite{Kotani2001, Ament2011, Groot2024, Mitrano2024}.
Its bulk sensitivity, combined with element and configuration selectivity, makes RIXS particularly well suited to study charge excitations and determine the energies of localized $f^n$ configurations and $U_{ff}$ \cite{Butorin1996, BUTORIN1999, Magnuson2001}, as well as spin excitations, $k_B T_K$ \cite{Hancock2018, rahn2022}, and the temperature dependence of the $4f$ occupation $n_f$ in Kondo systems. Furthermore, high-resolution RIXS can resolve low-energy excitations such as crystal-field and spin-orbit excitations, e.g.~in Ce compounds \cite{Amorese2016, Amorese2018, Amorese2018_2, Amorese2022, Amorese2023, Zhao2023}. 
Interestingly, RIXS exhibits a qualitatively different temperature dependence despite final states that are similar to those probed in PES and IPES. The intensity of the  $f^{0}$ structure decreases with increasing temperature, whereas the $f^{1}$ contribution increases and the $f^{2}$ intensity remains nearly constant. In this work, we demonstrate that this contrasting behavior arises from the different matrix elements that govern the RIXS process.

 Recently, we reported a RIXS study on Ce$_{0.93}$Sc$_{0.07}$ \cite{Tegomo2023}, a compound, which is well known to exhibit the $\gamma$-$\alpha$ first-order hysteretic isostructural transition \cite{zhu2020,wu2021}. We observed a systematic $T$-dependent hysteresis of the $f^0$ final-state spectral intensity upon cycling across the $\gamma$-$\alpha$ transition, and conjectured that this $T$-dependence reflects the Kondo energy scale. However, $T$-dependent RIXS measurements on a Kondo system without a first-order hysteretic transition are necessary to confirm this hypothesis.  CeSi$_2$ is considered a particularly suitable candidate for this investigation. It crystallizes in the tetragonal $\alpha$-ThSi$_2$ structure and exhibits the thermodynamic characteristics of a Kondo system. Its magnetic susceptibility shows Fermi-liquid behavior \cite{YASHIMA1982,YASHIMA1982_2, Lee1987, Shaheen1987} at low temperatures $T \leq 20$ K and follows a Curie-Weiss law between 200 K to 600 K. Heat capacity measurements exclude magnetic or crystal electric field transitions below 70 K down to 0.1 K and yield a linear coefficient of the specific heat $\gamma \approx \SI{100}{\milli\joule . \mole^{-1}.\kelvin^{-2}} $, which is strongly enhanced compared to a normal metal \cite{YASHIMA1982}. Magnetic susceptibility and specific heat measurements carried out on our sample batch are consistent with this behavior (see Supplemental Material (SM) \cite{SM}). These facts indicate a screening of the 4$f$ magnetic moment by conduction electrons at low-$T$ and that the Kondo effect dominates over the RKKY interaction. PES and IPES experiments showed $T$-dependent $f$-electron states close to the Fermi level, which is a signature of a Kondo system \cite{Pattey1987, Malterre1992, Grioni1997}.

In this Letter, we report the results of XAS and incident polarization-sensitive RIXS measurements of CeSi$_2$, which allow us to identify the $f^0$, $f^1\underline L^1$, and $f^2\underline L^2$ features in the spectra. The $T$-dependent measurements from 15 K to 300 K allow us to probe the spectral evolution across $T_{K}$. We use a simplified SIAM model \cite{Bickers1987} to calculate the $T$-dependence of the $f^0$ RIXS intensity, and then determine $T_K$, thus facilitating a Kondo scaling analysis. We further demonstrate that the thermal evolution observed in RIXS differs qualitatively from that reported in PES and IPES studies, reflecting the configuration-selective sensitivity of the different spectroscopic probes.

RIXS experiments were carried out at the PEAXIS spectrometer \cite{LIEUTENANT2016, Schulz2020} at BESSY II , and at the SEXTANTS \cite{Sacchi_2013,Sorin} beamline of synchrotron SOLEIL. The SIAM simulations were performed with the full multiplet \cite{SUNDERMANN2016, Sundermann2021, Chuang2021} using the QUANTY code \cite{Haverkort2012, Haverkort2016}.  Details of the experimental methods, the simulations, synthesis and characterization of the investigated samples are described in the SM \cite{SM}.

 \begin{figure}[htb]
\vspace{-0.2cm}
\begin{center}
\scalebox{0.7}{\includegraphics{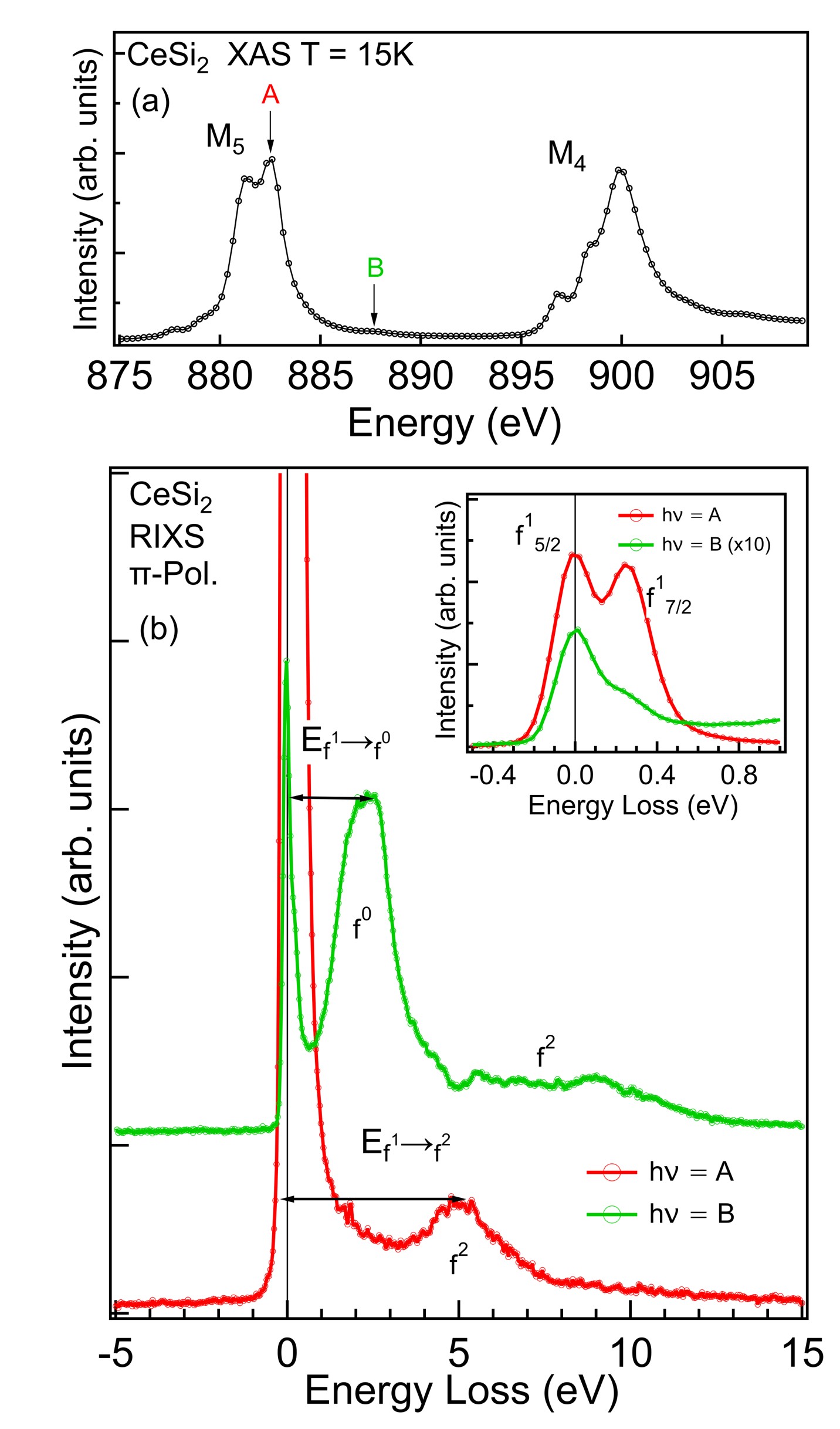}}
\end{center}\vspace{-0.5 cm}
\caption{\label{XAS_RIXS_CeSi2} (a) Ce $M_{4,5}$ edge XAS spectra of CeSi$_2$ measured at T = 15 K. (b) RIXS spectra of CeSi$_2$ taken with the incident photon energy tuned to $h \nu = \SI{882.5}{ \electronvolt}$ (label A in the XAS spectrum) and $h\nu  = \SI{887.5}{ \electronvolt}$ (label B in the XAS spectrum).} \vspace{-0.0 cm}
\end{figure}

\begin{figure*}[htb]
\vspace{-0.2cm}
\begin{center}
\scalebox{0.315}{\includegraphics{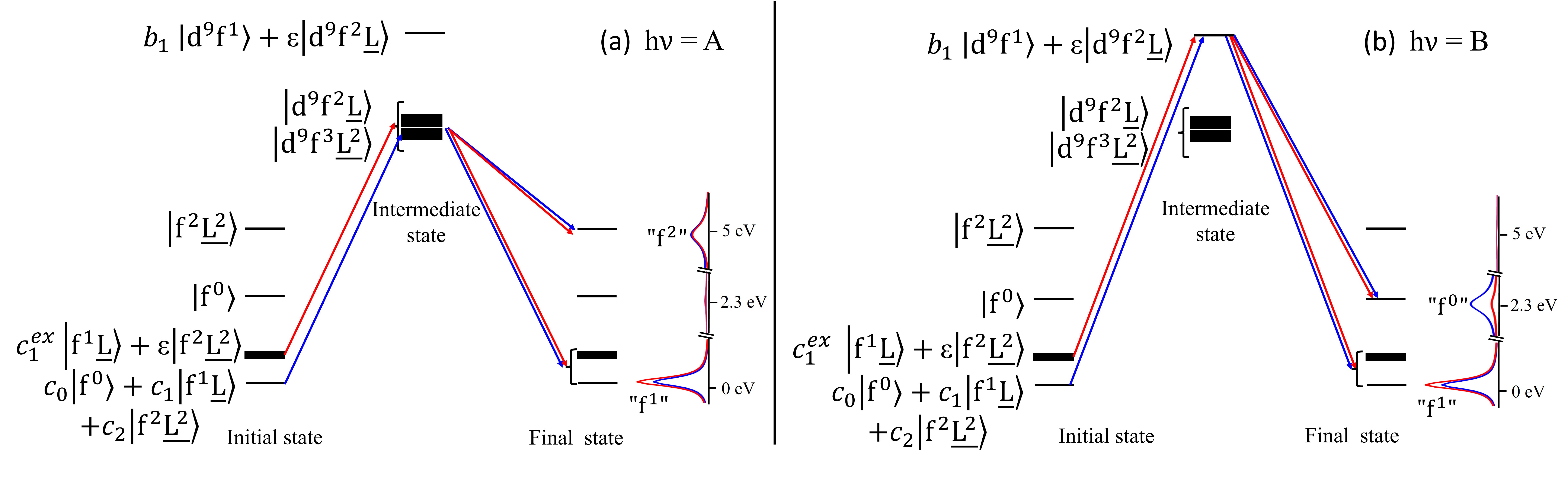}}
\end{center}\vspace{-0.5 cm}
\caption{\label{Energy_diagram} Schematic energy level diagram of a cerium Kondo system showing initial, intermediate, and final states of RIXS. The blue and red arrows represent transitions for $T = 0$ and $T\gg T_K$, respectively.  The blue and red schematic spectra represent spectra for $T = 0$ and $T\gg T_K$, respectively. 
For $h\nu =A$, at $T = 0$, the initial state is dominated by the ground state while for $T\gg T_K$ it is dominated by the first excited state. The arrows indicate transitions which explain the temperature dependence of $f^1$ and $f^2$ final states via both $d^{9}f^{2}$ and $d^{9}f^{3}$ in the intermediate state. Similarly, for $h\nu =B$ the arrows indicate transitions which explain the temperature dependence of $f^1$ and $f^0$ final states via the $d^{9}f^{1}$ in the intermediate state. Spin-orbit states are omitted for clarity.} \vspace{-0.0 cm}
\end{figure*}

 


Figure \ref{XAS_RIXS_CeSi2}(a) shows the Ce $M_{5}$ edge XAS spectra of CeSi$_2$  measured at T = 15 K in the total electron yield mode (TEY). The cleanliness of the sample was confirmed by the absence of the oxidation peak \cite{Yagci_1986} in the XAS spectra. The prominent structure centered at around 882 eV corresponds to the transitions 3d$^{10}$4f$^1$ $\rightarrow$ 3d$^{9}$4f$^2$ and 3d$^{10}$4f$^2$ $\rightarrow$ 3d$^{9}$4f$^3$, which overlap since the attractive Coulomb interaction $Q_{3d,4f}$ in the 3d$^{9}$4f$^2$ state with a 3d core hole is compensated by the repulsive Coulomb interaction $U_{ff}$ in the $f^3$ configuration \cite{Tegomo2022, Tegomo2023}. Moreover, the XAS spectrum exhibits a very small satellite at $h\nu$ = 887.5 eV, corresponding to the 3d$^{10}$4f$^0$ $\rightarrow$ 3d$^{9}$4f$^1$ transition. These three configurations confirm the intermediate valence character of the ground state \cite{Fuggle1983, Fuggle1983_2, Gunnarsson1983} which is expressed as \[\big|G\big\rangle = c_{0}\big|f^{0}\big\rangle + c_{1}\big|f^{1}_{5/2}\underline L\big\rangle + c_{1}'\big|f^{1}_{7/2}\underline L\big\rangle + c_{2}|f^{2}\underline L^2\big\rangle, \]
  where \underline L denotes a hole in the valence band. 
 The RIXS spectra obtained at $h\nu$ = 887.5 eV, corresponding to the tiny B feature in XAS, show a strong resonant enhancement of spectral features, enabling a precise quantitative analysis of the Kondo scenario.

\begin{figure}[b]
\vspace{-0.2cm}
\begin{center}
\scalebox{0.29}{\includegraphics{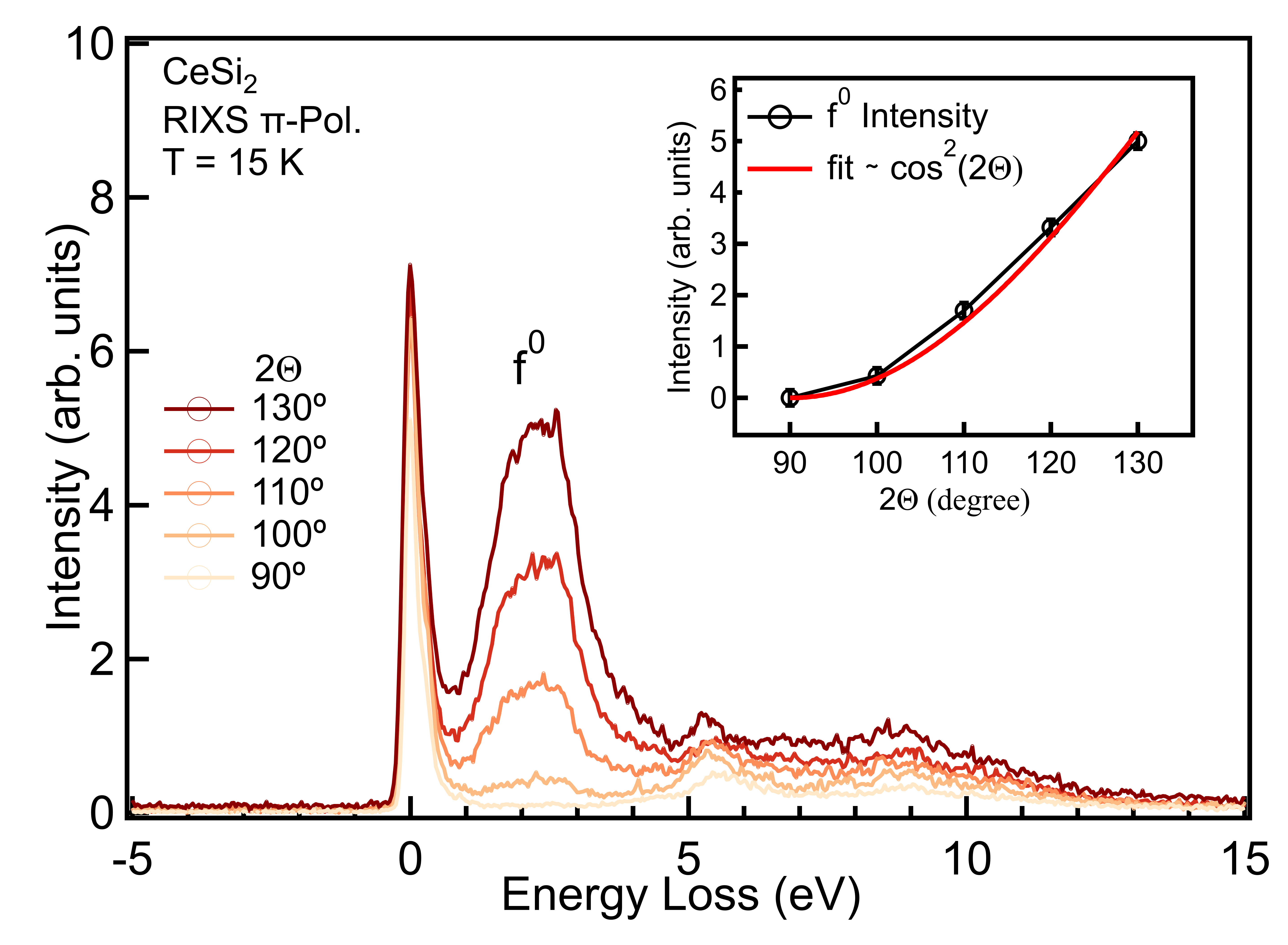}}
\end{center}\vspace{-0.5 cm}
\caption{\label{scattering_angle_dep} Scattering angle dependence of the CeSi$_2$ RIXS intensity at h$\nu$ = B with  $\pi$ polarization. The inset shows the evolution of the integrated $f^0$ intensity. } \vspace{-0.0 cm}
\end{figure}

Figure \ref{XAS_RIXS_CeSi2}(b)  shows the RIXS spectra of CeSi$_2$ measured at T = 15 K with the incident photon energies tuned on the XAS main structure (h$\nu$ = 882.5 eV = A) and on the satellite (h$\nu$ = 887.5 eV = B), at a scattering angle of $2\Theta = \SI{130}{\degree}$. The RIXS spectra for h$\nu$ = A and h$\nu$ = B exhibit a structure at $E_{f^{1}\rightarrow f^{2}}$ = 4.9 $\pm$ 0.2 eV and at $E_{f^{1}\rightarrow f^{0}}$ = 2.3 $\pm$ 0.1 eV, respectively. They correspond to a transition from the intermediate-valence ground state, with leading 4$f^1$ character, to a final state of mainly $f^2$ and $f^0$ character.
In the low-energy range, the structures at energy losses of 0 and 0.3 eV correspond to final states with $f^{1}_{5/2}$ and $f^{1}_{7/2}$ character, respectively. Following the calculations in the framework of the SIAM, the $f^{1}_{5/2}$ structure contains two contributions (if the crystal field effect is neglected): the transitions to the hybridized singlet ground state at $E = 0$ and the Kondo resonance at $\delta = k_{B}T_{K}$. However, only one peak is observed due to the limited energy resolution.

\begin{figure}[b]
\vspace{-0.2cm}
\begin{center}
\scalebox{0.28}{\includegraphics{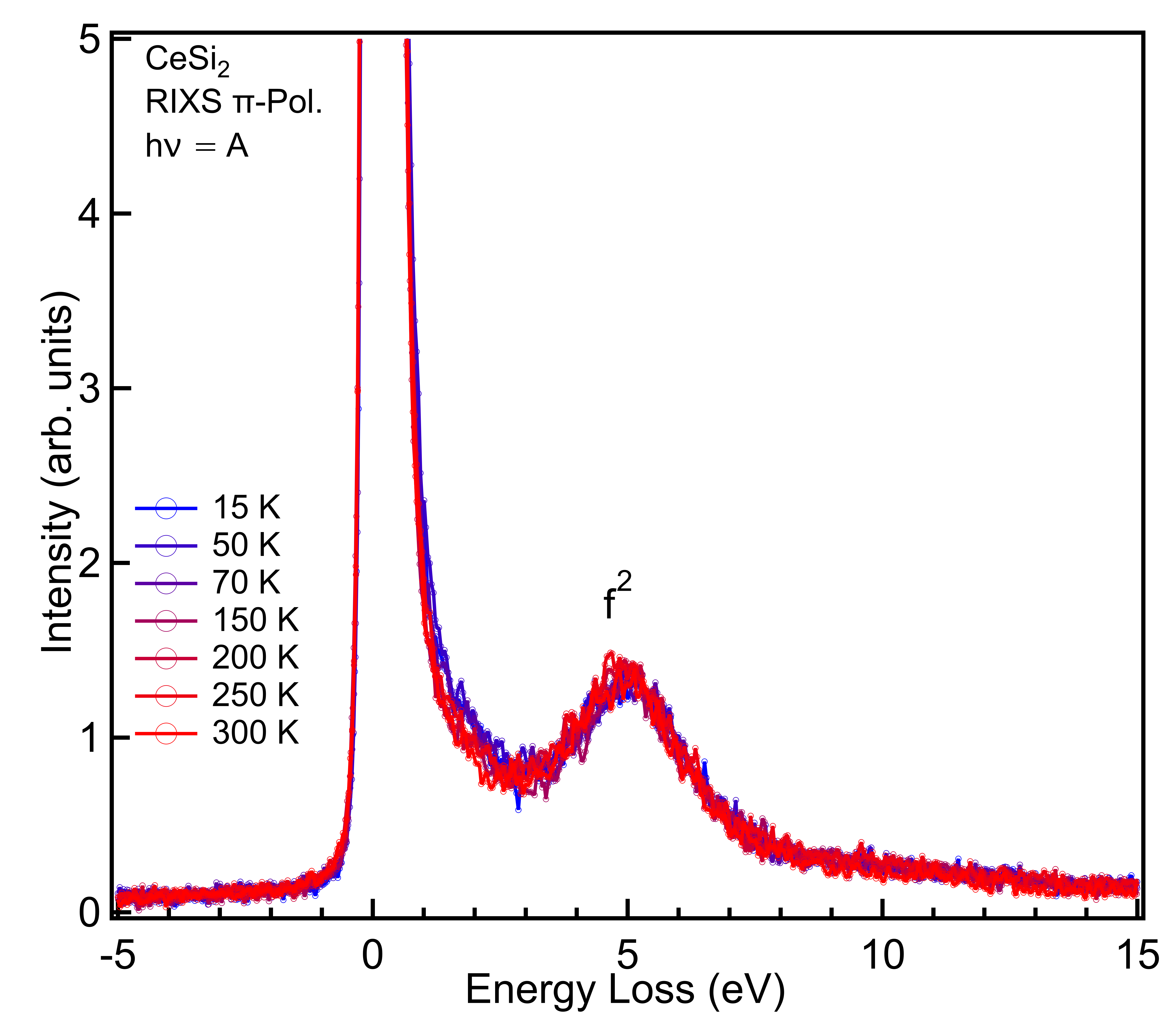}}
\end{center}\vspace{-0.5 cm}
\caption{\label{Tdep_res} Temperature dependent RIXS spectra of CeSi$_2$ with the incident photon energy  h$\nu$ = A. The $f^2$ final state is temperature independent. } \vspace{-0.0 cm}
\end{figure}

\begin{figure*}[htb]
\vspace{-0.2cm}
\begin{center}
\scalebox{0.55}{\includegraphics{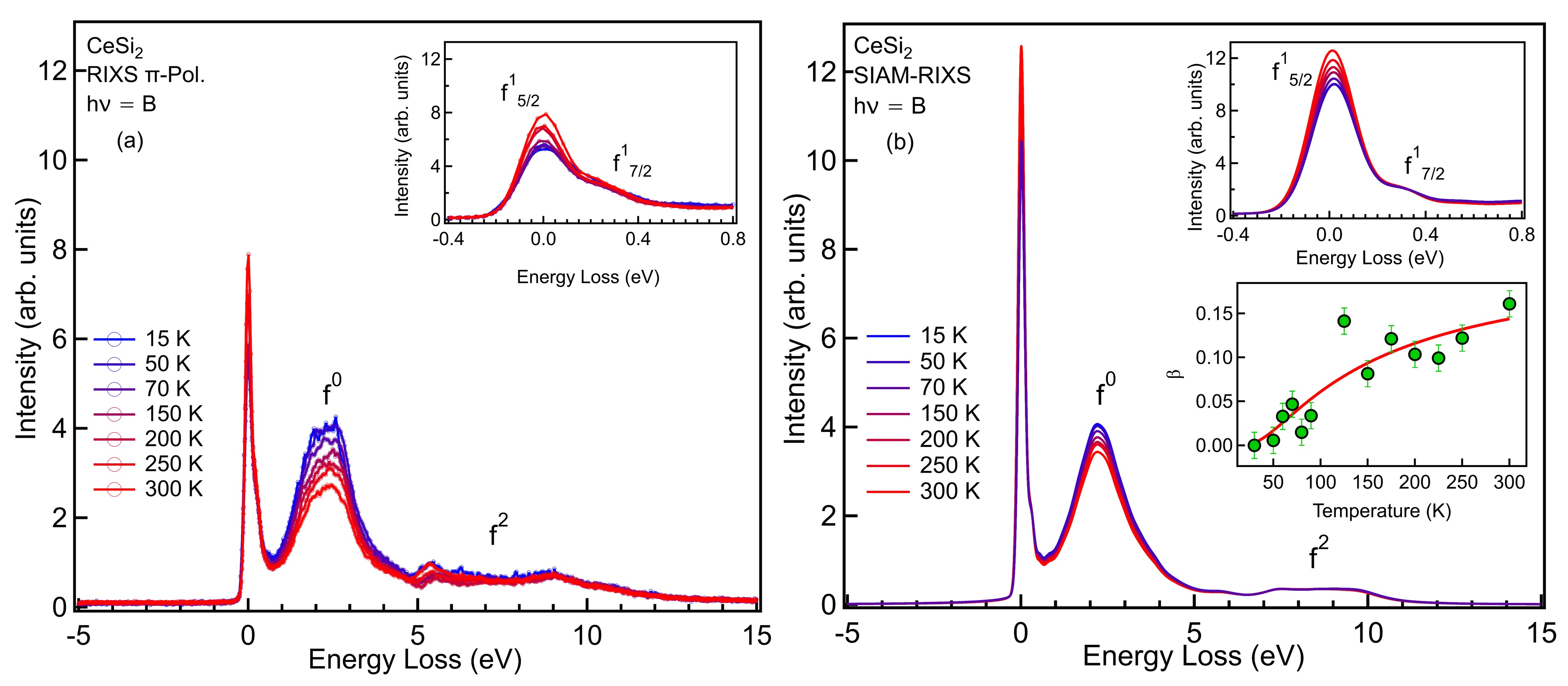}}
\end{center}\vspace{-0.5 cm}
\caption{\label{Tdep_f0_exp_siam} (a) Temperature ($T$) dependent RIXS spectra of CeSi$_2$ with the incident photon energy  h$\nu$ = B. A significant $T$-dependence of the $f^0$ final state is observed. (b) Calculated $T$-dependent RIXS spectra with $\pi$ polarization using the SIAM. The insets show an
enlarged view of the $f^1$ peaks and the evolution of the integrated $f^0$ intensity with temperature and its fitting using the Boltzmann weight of the local moment $\beta(T)$ (see \cite{SM}).} \vspace{-0.0 cm}
\end{figure*}

It is well known that the $f^0$ final state in RIXS spectra of cerium compounds exhibits a drastic polarization dependence \cite{Nakazawa2000, WATANABE2002, Sasabe2017, Tegomo2022, Sasabe2023, Tegomo2023}, where it is completely suppressed in $\pi$ polarization for \SI{90}{\degree} scattering angle. Although it has been theoretically shown that $f^0$ is allowed in $\pi$ polarization at scattering angles other than \SI{90}{\degree}, this has never been experimentally observed. It is possible to decompose the RIXS scattering cross section into linear combinations of fundamental processes in which an angular momentum $J=0,1,2$ is coupled to the sample, with coefficients given by the polarization geometry \cite{Ament2011, Myrtille, Michelangelo}. In this framework, selection rules for the probed excitations with weight $\chi_{(J)}$ corresponding to angular momentum $J$ apply as $J_{fs}\in\{J_{gs}\otimes J_{\chi}\}$, with $J_{gs}$ and $J_{fs}$ the total angular momenta of the ground and final states, respectively. 

Therefore, to further confirm that the structure at 2.3 eV corresponds to the $f^0$ final state, we have measured the RIXS spectra as a function of the scattering angle as shown in Figure \ref{scattering_angle_dep}, here in a $\pi$ scattering geometry with no polarization filter on the scattered radiation. The total intensity for incident $\pi$ polarization is:
\begin{align}
    I_{\boldsymbol{\hat{\pi}}_{\mathrm{in}}} &= \frac{1}{6}\left(\chi_{(0)} + 2 \chi_{(2)}\right)\cos^2(2\Theta) \nonumber \\ 
    & \quad + \frac{1}{4}\left(\chi_{(1)} +\chi_{(2)} \right) \left( 1 + \sin^2(2\Theta) \right). 
\label{PiAngDependentScattering}
\end{align}
The experimentally observed $f^0$ scattering angle dependence for $\pi$ polarization follows $\cos^2(2\Theta)$ and is thus associated with the $\chi_{(0)}$ scattering contribution. In the ground state, the $f^0$ configuration has a total angular momentum $J=0$. The $f^{1}_{5/2}\underline L$ 
 ($f^{1}_{7/2}\underline L$) configuration consists of a 4$f$ electron with angular momentum $j=5/2$ ($j=7/2$) and a hole in the conduction band states with angular momentum $j_{\underline L}=5/2$ ($j_{\underline L}=7/2$). So, the total angular momenta of the  $f^{1}_{5/2}\underline L$ 
 ($f^{1}_{7/2}\underline L$) configuration are $J = 0,1,2, ...,5$ ($J= 0,1,2, ...,7$) since 
$|j-j_{\underline L}|\leq J \leq j+j_{\underline L}$. Given that the hybridization term of the SIAM conserves the total angular momentum $J$, only its matrix elements between states with the same total angular momentum make a finite contribution. Therefore, only the states $f^{1}_{5/2}\underline L$ and $f^{1}_{7/2}\underline L$ with $J=0$ hybridize with the $f^0$ ($J=0$) states and give rise to the singlet ground state with total angular momentum $J=0$. The remaining non-hybridized states with $J=1,2, ...,5$ and $1,2, ...,7$ are the low-lying excited states (see Fig. \ref{Energy_diagram}). 

The $T$-dependent RIXS spectra of CeSi$_2$ measured with incident photon energies h$\nu$ = A and h$\nu$ = B are shown in Figures \ref{Tdep_res} and \ref{Tdep_f0_exp_siam}(a), respectively. For h$\nu$ = A (Figure \ref{Tdep_res}), the $f^2$ feature remains essentially unchanged on changing $T$, whereas for h$\nu$ = B (Figure \ref{Tdep_f0_exp_siam}a), the $f^0$ structure is strongly depleted on increasing temperature. Simultaneously, the intensity of the $f^{1}_{5/2}$ final state increases on increasing $T$ (Figure \ref{Tdep_f0_exp_siam}a inset).

A comparison of the RIXS results with well-known PES and IPES results reveals the contrasting $T$-dependent behavior as probed by each technique and originating from the different 4f character states ($f^0$, $f^1$ or $f^2$). In PES, the structure associated with final states with mainly $f^0$ character is $T$-independent \cite{Malterre1996}, whereas in RIXS, its intensity decreases significantly as $T$ increases. Conversely, the structure associated with final states mainly of $f^2$ character in RIXS shows negligible $T$-dependence, whereas in IPES its intensity increases with $T$ \cite{Malterre1992,Malterre1992_2}.

Since $T_K$ plays the role of a characteristic temperature scale, it allows us to qualitatively understand the observed $T$-dependence. In fact, the hybridized singlet ground state is separated from the low-lying first excited magnetic state by the Kondo energy $k_B$$T_K$. Following the above-mentioned selection rules, the $\chi_{(0)}$ scattering process leaves the angular momentum of the ground state unchanged, such that coupling to $J_{fs}=0$ is only possible starting from the Kondo-singlet state with $J_{gs}=0$ and forbidden from local moment states.  As a consequence, the $f^0$ intensities scale with the thermal weight of the Kondo-singlet in the system ground state, i.e., it scales with $T/T_K$.

Along with symmetry selection rules, the ground-state configuration occupancies also play a role for the spectral intensities. In the Ce $M_5$ RIXS process at low-$T$ ($T$ $\ll$ $T_K$), only the singlet ground state is occupied, and the final states with $f^1$, $f^0$ and $f^2$ character are reached (see Figure \ref{Energy_diagram}), reflecting the amplitude of the $f^1$, $f^0$ and $f^2$ configurations in the ground state. The $T$-dependence can be qualitatively understood within a simple physical picture. With increasing $T$, the low-lying excited states given by: $
 \big|\Psi_{ex}\big\rangle = c_{1}^{ex}\big|f^{1}_{5/2}\underline L\big\rangle + c_{1}^{',ex}\big|f^{1}_{7/2}\underline L\big\rangle + c_{2}^{ex}\big|f^{2}\underline L^2\big\rangle $) (which contain no $f^0$ character), are progressively populated. These excited states primarily couple to the final states with $f^{1}$ and $f^{2}$  character (see Figure \ref{Energy_diagram}), causing the $f^0$ RIXS spectral weight to decrease with increasing $T$ following the evolution of $1-n_f$. Since the $f^2$ weight in the low-lying excited state $c_2^{\rm ex}$ is comparable to that in the ground state $c_2$ ($c_2^{\rm ex}$ $\approx$ $c_2$), the $f^2$ RIXS spectral weight remains essentially unchanged with increasing $T$. 
 What about the $f^{1}_{5/2}$ structure? At $T = 0$, only the Kondo singlet ground state is occupied, and the $f^{1}_{5/2}$ feature arises from transitions to two possible states (see Figure \ref{Energy_diagram}): (i) the Kondo singlet state and (ii) the low-lying excited magnetic states, with respective weights $(c_1)^2$ and $c_1 c_1^{\rm ex}$. In the Kondo limit ($n_f \approx 1$), $c_1$ is close to unity, while $c_0$ and $c_2^{\rm ex}$ are very small, and $c_1^{\rm ex} > c_1$. As $T$ increases, the low-lying excited states get progressively populated. Because these states have predominantly $f^1$ character, they also couple to the singlet and magnetic final states, with weights $c_1 c_1^{\rm ex}$ and $(c_1^{\rm ex})^2$, respectively. Since $c_1^{\rm ex} > c_1$, the total integrated spectral weight of the $f^{1}_{5/2}$ structure increases with increasing $T$.
 
 The $T$-dependence of the $f^0$ RIXS final state is consistent with the SIAM calculation \cite{Bickers1987}, which predicts a similar evolution for the Kondo resonance above the Fermi level. To confirm this agreement, we performed a $T$-dependent calculation of the RIXS spectra using the SIAM, a well-established theoretical method that yields both dynamical and static properties. 
 We performed the calculation with $n_{f}(0)$ = 0.97 and using a crystal electric field (CEF) ground-state doublet and two excited CEF doublet states at $\Delta_1$ = 25 meV and $\Delta_2$ = 48 meV \cite{GALERA1989}. Figure \ref{Tdep_f0_exp_siam}(b) shows the $T$-dependent calculated spectra.  The best agreement (see Supplemental Material \cite{SM} for more details) with experiment and theory is obtained for a Kondo temperature $T_{K} =  129 \pm 32 $ K. This value of $T_{K}$ is higher than the value estimated from ultraviolet PES and 4d-4f resonant PES (RPES) studies \cite{Pattey1987, Garnier1997, Lawrence1993}; however, it is consistent with the $T_{K}$ determined from the specific heat \cite{YASHIMA1982_2} as well as 3d-4f RPES  \cite{CHOI2004} measurements. The difference of $T_K$ between the present study and low-energy PES studies is likely due to the strong surface contribution in low-energy PES studies.


In summary, we have employed RIXS at the Ce $M_5$ edge to investigate the $T$-dependence of charge and spin excitations in the heavy-fermion compound CeSi$_2$. The measurements reveal a pronounced $T$-dependence of the $f^0$ final state, which has never been observed before. It provides direct spectroscopic evidence of a Kondo energy scale in a heavy-fermion system. In addition, the thermal evolution observed in RIXS contrasts sharply with previous PES and IPES studies, reflecting the configuration-selective sensitivity of the different probes. In particular, the $f^0$ feature is $T$-independent in PES but shows a clear $T$-dependence in RIXS. In contrast, the $f^2$ feature is $T$-dependent in IPES, but $T$-independent in RIXS.
The observed experimental results are well described by the single-impurity Anderson model, and are establishing RIXS as a reliable bulk-sensitive probe of strongly correlated heavy-fermion systems.


\textit{Acknowledgments}-We thank the Helmholtz-Zentrum Berlin für Materialien und Energie (Berlin, Germany) for the allocation of synchrotron radiation beamtime under proposal N° 242-12684. We acknowledge the French Synchrotron facility SOLEIL (Saint-Aubin, France) for the allocated beam time under proposal N° 20240502. We acknowledge access to the CoreLab Quantum Materials at Helmholtz-Zentrum Berlin für Materialien und Energie, HZB, where the magnetic susceptibility and heat-capacity measurements were performed. This work was supported by a public grant overseen by the French National Research Agency (ANR) as part of the BONASPES project (ANR-19-CE30-0007).

\bibliography{Ref.bib}

\end{document}